\documentclass[pre,twocolumn,showpacs,superscriptaddress]{revtex4}

\usepackage{graphicx}
\usepackage{amsmath,amsfonts,amssymb}

\begin{document}

\title{Local Symmetries and Order-Disorder Transitions in Small Macroscopic Wigner Islands}
\date{\today}
\author{Gwennou Coupier}\affiliation{Laboratoire Mati\`ere et Syst\`emes Complexes, Universit\'e Paris 7, Unit\'e mixte du C.N.R.S 7057, 140 rue de Lourmel, 75015 Paris, France}
\author{Claudine Guthmann}\affiliation{Laboratoire Mati\`ere et Syst\`emes Complexes, Universit\'e Paris 7, Unit\'e mixte du C.N.R.S 7057, 140 rue de Lourmel, 75015 Paris, France}
\author{Yves Noat}\affiliation{Institut des NanoSciences de Paris, Universit\'es Paris 6/Paris 7, Unit\'e mixte du C.N.R.S 7588, 140 rue de Lourmel, 75015 Paris, France}\author{Michel Saint Jean}
\email{saintjean@gps.jussieu.fr}\affiliation{Laboratoire Mati\`ere et Syst\`emes Complexes, Universit\'e Paris 7, Unit\'e mixte du C.N.R.S 7057, 140 rue de Lourmel, 75015 Paris, France}

\begin{abstract}
The influence of local order on the disordering scenario of small Wigner islands is discussed. A first disordering step is
 put in evidence by the time correlation functions and is linked to individual excitations resulting in configuration transitions,
 which are very sensitive to the local symmetries. This is followed by two other transitions, corresponding to orthoradial and radial
diffusion, for which both individual and collective excitations play a significant role. Finally, we show that, contrary to large systems,
the focus that is commonly made on collective excitations for such small systems through the Lindemann criterion has to be made carefully in order to clearly identify
 the relative contributions in the whole disordering process.
\end{abstract}

\maketitle

Many efforts have been intended in past years to understand properties of mesoscopic devices in which interacting particles are confined.
For instance, these particles can be vortices in mesoscopic shaped superconductors \cite{blatter94}\cite{hata03}, electrons in quantum dots \cite{ashoori96}, strongly coupled rf dusty
plasma \cite{liu99}, trapped cooled ions \cite{gilbert88}, vortices in superfluid $He4$ \cite{yarmchuk82}, electron dimples on a liquid helium surface \cite{leiderer82}, vortices
 in a Bose-Einstein condensate \cite{chevy00} or colloidal particles \cite{zahn99}.
More recently, in order to take advantage of the macroscopic scale to explore the properties of such systems, we have proposed a new macroscopic system
consisting of $N$ interacting charged balls of millimetric size free to move on a plane conductor and confined electrostatically, the temperature being simulated
by a mechanical shaking  \cite{stjean04}. Using this system we observed the equilibrium configurations of the Wigner islands obtained for circular \cite{stjean01}
and elliptic confinements \cite{stjean02}.

Our previous study, essentially focused on static properties,
conclude that at low temperature and for a circular confining potential, the observed small islands $(N<40)$
present self-organized patterns constituted by concentric shells in which the balls are located. As it was widely discussed in the
literature \cite{campbell79}\cite{bedanov94}\cite{lai99},  this peculiar structure is due to the competition between the ordering into a triangular
lattice symmetry, which appears for infinite two-dimensional electrostatic systems, and the circular symmetry imposed by the confining potential. In the following, the Wigner
islands will be described by the configuration $(N_0-N_1-N_2-..)$ where $N_i$ is the number of balls in the i\textsuperscript{th} shell from the center.

Surprisingly, the influence of the thermal fluctuations on the phase behavior which was an important question largely studied for the
 two-dimensional extended systems has been little discussed in such small systems. The studies devoted to this question are essentially numerical and focused on
 the temporal stabilities of the different configurations \cite{schweigert98_1}\cite{kong02}, their spectral properties \cite{schweigert95}\cite{partoens04} or the mean
displacements of the particles \cite{bedanov94}.
 Let us however indicate an interesting experimental work concerning the``melting" process of
colloidal interacting particles islands \cite{bubeck99}\cite{bubeck02}.``Melting" in two-dimensional quantum electron clusters has also been recently studied in \cite{filinov01}.

Phase transitions in two-dimensional large crystals have mainly been described by the changes in the asymptotic behavior of spatial correlation functions.
For instance, the Kosterlitz-Thouless-Halperin-Nelson-Young (KTHNY)  theory
predicts a two-step melting scenario according to which the liquid phase is reached  when bond-orientational correlations become short-range \cite{young79}\cite{nelson79}.
 Parallely, it has been suggested that
temporal correlations should have the same behavior as the spatial ones \cite{nelson83}. This has recently been put in evidence experimentally for colloidal
 systems \cite{zahn99}\cite{zahn00}.

On the other hand, the previous studies on small interacting systems always refer to a generalization of the well-known Lindemann's method employed
 to describe the order-disorder transition for large systems  \cite{bedanov85_2}, which considers the liquid phase is reached when the mean square
 displacements relatively to the lattice parameter of
 the particles go beyond the value $\gamma_M=0.1$ (which seems to be independent
of the interaction  \cite{bedanov85_2}\cite{lozovik85_3}), or 0.05 for each coordinate. Note that it has been shown
 that, for infinite systems, the transition temperature exhibited there is the same as the one given by the correlation functions \cite{bedanov85_1}. Let us underline that
no such result is known for small systems.

For the latter, the studies lying on Lindemann's criterion predict or observe that the shell-structured islands become less ordered while the temperature increases.
They describe a two-step process corresponding to two different transition temperatures. At very low temperature, each particle
 is thermally excited in its local potential; a first transition appears at the temperature $T_O$ when the orientational order between the shells is lost.
This first transition is followed by a second one, at the temperature $T_R$, which corresponds to the emergence
 of the radial diffusion of particles between the shells, as well as an angular diffusion in the shells. For higher temperature, the initial order is completely destroyed.
The transition temperatures $T_O$ and $T_R$ are respectively identified as the temperatures at which the intershell angular and radial mean square displacements
 have a rapid and strong increase \cite{bedanov94}.

In this paper, we discuss the influence of the local order on the ``melting''. Indeed, for small islands, singular events such as a unique jump of only one particle from one shell to another one
 result to an important
 modification of the system since such an intershell jump corresponds to a transition between well-separated stable and metastable equilibrium
configurations of different geometry. Consequently, such jumps will modify the bond-orientational correlation functions. We shall show that,
for a given temperature, those configurations switches and thus the disordering process are controlled by the local geometry. In order to distinguish between those events
and the collective excitations of the particles, we calculate the Lindemann parameters on each shell and for each configuration separately, without taking
 into account jumps between shells. We determine the temperatures
 at which the collective excitations appear and destroy the ordered configurations. This unusual Lindemann procedure is in agreement
with our definition of an ordered system as a system always close to its most symmetrical configuration. For instance, an intershell jump which
 does not destroy the local symmetry does not have to be considered as relevant in our description. This will explain why the obtained critical temperatures will be higher than
the one determined by the usual Lindemann method which includes all kind of displacements in the calculation of the parameters, as in \cite{bedanov94}.
 Notice that this question is specific to small systems since in large systems there is
a continuum of states and individual excitations are hidden by collective ones.

In order to evaluate the relative importance of these different contributions to the ``melting'', we have experimentally observed the evolution of macroscopic Wigner crystals
while the effective temperature is increased. To emphasize the contribution of the individual excitations with respect to the collective
ones, we have selected a set of systems which have very different local symmetry for a similar number of balls in order to present
different configuration transition behaviors for almost the same kind of collective excitations.
We chose Wigner islands consisting in $N=18,19$ and 20 interacting particles confined in a circular frame. These systems, in spite
of a very close number of balls, are very different from the local symmetry point of view and the resulting excitation energy spectra. Indeed, the ``magic number"
system $N=19$ exhibits a three fold symmetry, as its ground configuration is (1-6-12). In fact, the latter shell can be divided into two
 subshells of 6 balls, as it was numerically shown in \cite{campbell79}, however, the difference between the two radii being rapidly of the same order as the thermally induced
radial fluctuations, we will still refer to this configuration as the (1-6-12) one. By contrast, the ground configurations of $N = 18$ and $N = 20$ systems are constituted
 by incommensurable shells (in the sense that the ratio between the number of balls on each shell is not an integer), their ground states being respectively (1-6-11) and (1-6-13).
When the temperature increases, the two lower excited states of each system can be reached. In spite of the radial displacements due to the temperature, the ring-like structure
 remains and the configurations can be very well identified :
\begin{itemize}
\item For 18 balls : (1-6-11), (1-5-12), (0-6-12)
\item For 19 balls : (1-6-12), (1-7-11), (1-5-13)
\item For 20 balls : (1-6-13), (1-7-12), (2-6-12)
\end{itemize}
We can notice that only one ball jump from a shell to another is necessary to induce a configuration switch.

These configurations correspond exactly to those computed in \cite{campbell79} for logarithmic interparticle interaction potential, strongly
suggesting this kind of interaction between the balls, at least  within the range of our experimental inter-particle distance \cite{stjean01}. This conclusion has been confirmed
 later by comparing the ground configuration obtained for elliptic confinement with the configuration of vortices calculated in similar shaped mesoscopic superconductors
for which the inter-vortices interaction is logarithmic in the considered range of inter-particle distance \cite{stjean02} \cite{meyers00}.
From the energetic point of view, the local symmetry differences between the various systems induce strong differences in their excitation energy spectra \cite{campbell79}. The
configuration energies for $N=19$ and $N = 18$ are well separated, however the gap between the ground and the first excited state is larger for the magic number system.
By contrast, the latter is very small for $N = 20$.

In section \ref{procedure}, we present the experiments which validate the mechanical shaking as an effective thermodynamic temperature and we describe
 the parameters used to characterize the disorder, the configuration transitions and the collective excitations.
The evolutions with temperature of these parameters for $N = 18, 19$ and 20 islands will be described and
 discussed in details in section \ref{transition}. In section \ref{order}, the respective influence on the disordering of the individual and collective excitations will be discussed.
We will show that, according to the considered parameter, different transition temperatures can be identified. In particular,
the systems present an important configuration transition activity inducing disorder at a temperature smaller than those characterized by the Lindemann criterion. The ``melting'' will have
to be considered rather as a disordering than a real melting.

\section{\label{procedure}Experimental procedures and characterization parameters}

Our Wigner islands are constituted by millimetric stainless steel balls (of diameter $d = 0.8$ mm and weight $m=2.15$ mg) located on the bottom
 electrode of a horizontal plane capacitor (a doped silicon wafer whereas the top electrode is a transparent conducting glass). An isolated metallic circular frame
  of diameter $D =10$ mm and height $h=1.5$ mm intercalated between the two electrodes confines the balls  \cite{stjean01}. When a potential $V$ is applied to
the top electrode (the bottom one and the frame
being linked to the ground), the balls become monodispersely charged, repel each other and spread throughout the whole available space.
For the currently used potential $V$ (around a few hundred of volts), the charge of each ball has been evaluated to about $10^9$ electrons.
\begin{figure*}
\resizebox{1.9\columnwidth}{!}{\includegraphics{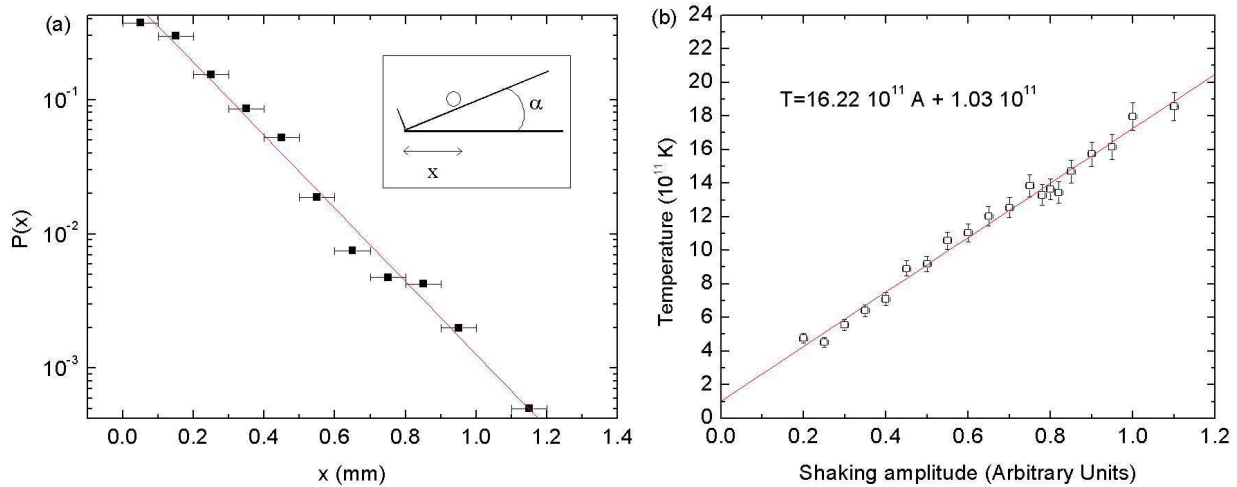}}
\caption{\label{temperature} Temperature calibration : (a) $x$ position distribution for a trapped ball moving on an inclinated plane (inset) and submitted to mechanical shaking (log scale).
The good fit with Boltzmann law validates the mechanical shaking as an effective
temperature. (b) Calibration of the effective temperature. }
\end{figure*}
The whole cell is fixed on a plate linked with two independent loudspeakers supplied by a white noise voltage. As we shall show later,
 this horizontal shaking results in an erratic movement of the balls which simulates an effective temperature that can be modified by tuning the shaking
amplitude. In the experiment presented here, the system is always prepared in its ground configuration at low effective temperature and the temperature is
progressively increased.

 Throughout the experiment, images of the arrays of balls are recorded in real-time using a CCD camera, and their center of mass is detected.
For each temperature, the positions of each ball are followed during 400 s, the minimum time between two snapshots being 100 ms. The characteristic time of the oscillation of a single ball in its
 local potential is, at low temperature, about half a second. Then, the choice of the total experiment time and the time interval allow us to track the balls (at least in the
 considered temperature range) and get statistically relevant data.
 The different configurations reached are then determined by counting the number of balls in each shell. In this procedure the radial
 limits of a shell are defined as the mean values between its radius and its neighbor's one, that have been measured in the ground state. They are independent from the temperature.

\subsection{Temperature calibration.}

Before studying the ``melting'' of such systems, strong attention has been paid to show that the cell shaking effectively results in a
 brownian motion of the balls allowing the identification of this shaking with an effective temperature. We present here the experiments
 performed in order to calibrate this effective temperature and the ``in situ" thermometer we have developed.

As in \cite{pouligny90}, we used a system for which the energy is well-known : the calibration was obtained by the
 use of a single ball rolling on the silicon wafer which has been winded with an angle $\alpha=25'$ from the horizontal plane. The ball can elastically bounce
 on a bottom wall. No electrostatic force is at stake and the only energy is the gravitational potential. For each voltage $A$ applied on the loudspeakers, the horizontal distances $x$ of the ball
 center from the bottom wall
 have been measured through the capture of a few thousand snapshots of the ball. The recorded random $x$ positions
 are distributed following the density $P(x)$.

 In order to compare this density with Bolzmann theory, it was fitted
 with the function $$P(x)=\frac{m g  \tan \alpha}{k_B T}e^{\frac{- E(x)}{k_B T}} $$
 where $E(x)= m g x \tan \alpha $ is the potential energy of the
 ball and $T$ the fit parameter which corresponds to the expected effective temperature, $k_B$ being the Boltzmann constant.

Figure \ref{temperature}(a) presents the experimental
data obtained for the voltage amplitude $A=1.0$ a.u. and the corresponding fit. At evidence Boltzmann law is obeyed. This very good agreement being observed whatever the voltages
 $A$, we may conclude that the mechanical shaking corresponds to an effective temperature. This effective temperature is obtained thanks to the fitting analysis, or more simply
 through the relationship $T={m g <x>\tan \alpha}/{k_B }$ and a calibration curve $T(A)$ has then been be obtained. As shown in figure \ref{temperature}(b), the relation
 between $A$ and $T$ is
affine within the range required to study the ``melting". Let us indicate that the temperature range is about $10^{11}$K, which has no other signification than the energy range.

\begin{figure}
\resizebox{0.95\columnwidth}{!}{\includegraphics{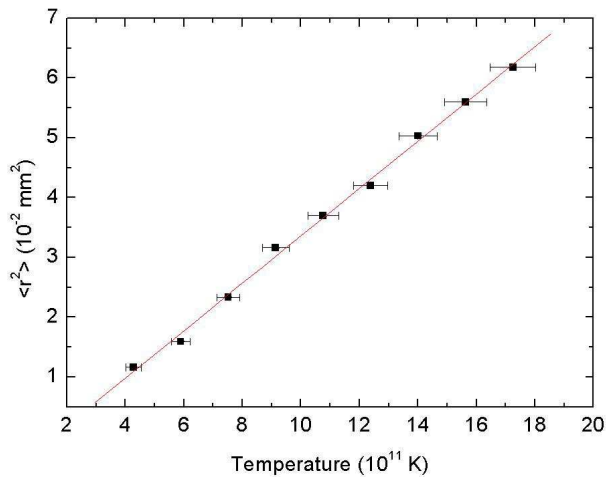}}
\caption{\label{excursionbilleseule} Mean square displacement $<r^2>$ of the ball plotted versus temperature for $D$=6 mm and  $V$=900 V. }
\end{figure}

 Finally, we have used this calibration to develop an``in-situ" thermometer which is constituted by a single ball
 trapped in a second circular frame located near the main one and submitted to the same voltage. The effective temperature is determined by measuring the radial mean square displacement $<r^2>$
of this unique ball for each given shaking amplitude $A$ and by identifying this displacement with the temperature T(A) previously determined by the calibration procedure. Whatever the
various thermometer diameters tested in order to optimize the sensibility of this thermometer ($D =5,6,7,8$ and 10 mm) and the applied potential $V$ ($V=700,800,900, 1000$ V) the mean
 square displacement $<r^2>$ varies linearly with the temperature. Same linearity is observed for the mean square speed. Figure \ref{excursionbilleseule} presents
 the variation $<r^2>$ (T) corresponding to the potential $V=900$ V
and the retained diameter $D=6$ mm. This thermometer will give a  precise determination of the temperature for any experiment to come, independently from
the variations associated to the total weight of the support or to the loudspeakers' ageing.

\subsection{Correlation functions}

The bond-orientational correlation function $g_6(t)$ is relevant in order to characterize phase transitions in two-dimensional systems. It is defined by $$g_6(t)=|<e^{i 6 (\theta(t+t_0)-\theta(t_0))}>|$$ where
$\theta(t)$ is the angle of a fixed bond between two particles and $<>$ denotes an averaging over all bonds. In infinite
 two-dimensional systems, $g_6$ tends to a constant roughly equal to 1 at low temperature. The system is then like an ordered crystal.
 If the temperature is higher than a temperature $T_{l}$, a strong decay with time
is observed, which denotes a liquid phase.

Such a dynamical
criterion offers experimental facilities since it is often easier to have long time acquisition rather than to observe a large sample of particles. In particular, we can readily use this criterion to
characterize order-disorder transitions in small systems.

Note that the factor 6 in the exponential is adapted for hexagonal lattices,
therefore we can expect the limit value at low temperature to be lower than 1 for our small systems in which the three-fold symmetry is broken. On the other hand,
 the averaging being made only on $N(N-1)/2$ links, we can't expect $g_6$ to go until zero value in the liquid phase, even at large time.
These points apart, this parameter still measures the correlations and we can expect to measure the same global behaviors according to the temperature.

\subsection{Configurations transition parameters}

Two parameters easy to get experimentally have been identified in order to characterize the configuration transitions.
The first of them is the jump rate $R_J$ :

$$R_J=\lim_{t\to\infty}\frac{N_J(t)}{t}$$

where $N_J(t)$ is the total number of configuration switches during the period $t$.
This parameter is an indicator of the ``transition activity" of the system. Qualitatively, this
 parameter is small at low temperature when only a few transitions occurs and increases strongly with temperature
when energetical barriers can be overcome.

A second way to characterize more quantitatively the transition rate is to measure the mean time required to escape from
each state or the mean residence time in each state. Let us consider for instance a Wigner island at low  temperature ;
this system can be understood as a two level system characterized by a  ground state $E_g$ and one metastable state $E_m$ (the
 only one that is reachable if the temperature is sufficiently low).
 The thermal fluctuations
 induce transitions between those two configurations. These transitions in the real space can be mapped in the phase space by the
 jump of a fictive particle from a well to another, characterized by their depth $E_g$ and $E_m$ and a saddle point $E_s$.

Under the assumption that all escape attempts are independent and of weak probability, the probability for the particle to stay in a state of energy $E$ during a time $\tau$ is :

$$P(\tau)=\frac{1}{\tau_0}e^{-\tau/\tau_0}$$

where $\tau_0=<\tau>$ is the mean residence or Kramers time. It depends on the energy barrier and on the temperature and is given by \cite{kramers40} :

$$ <\tau> = \tau_r e^{(E_s-E)/k_B T}$$

where $E_s-E$ is the barrier energy, and $\tau_r$ is the relaxation time within the well.

The energetical barriers can be measured by the slope of the curve describing the variations of the mean residence times in log scale as a function of 1/T. Similarly,
 the spectrum of the excitation energies is determined through the ratio of mean residence times in the two wells.

Qualitatively, this analysis can be extended to the case of high temperature for which the system explores more than the two first levels and reaches higher
 excited levels.  But the possibility for a particle to escape a well in different ways whose relative weight should depends on the geometry leads to a more acute problem.

We have tested the validity of this analysis in the case of the two-level systems $N=5$ and $N=6$. These two cases are interesting  since they
involve in their respective configuration transitions the two kinds of individual jump observed in the Wigner excitations. For $N=5$, the ground configuration consists in a unique
 shell (5) whereas the metastable configuration (1-4) requires a centripetal displacement of a ball. By contrast, the $N=6$ ground state is a centered configuration (1-5) and the
 metastable configuration (6) is reached after a centrifugal displacement of ball.

 \begin{figure}

\resizebox{0.95\columnwidth}{!}{\includegraphics{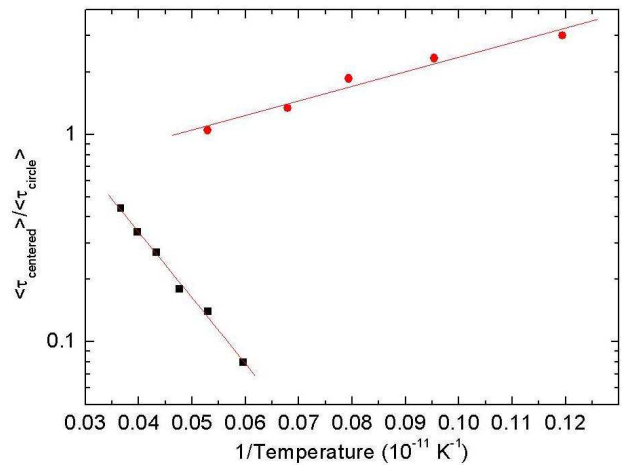}}
\caption{\label{temps56} Evolution with the temperature of the residence mean times ratio for two-level systems : 5 balls ($\blacksquare$) and 6 balls ($\bullet$). Ratios are in log
scale. Note the signs of the slopes that prove that (5) and (1-5) are the ground states for the 5-ball and 6-ball systems respectively.}

\end{figure}

  In figure \ref{temps56} we present the
 variation with the temperature of the ratio $<\tau_{centered}>/<\tau_{circle}>$, where $\tau_{centered}$ stands for the residence time in the centered configuration, respectively (1-4) or (1-5),
and $\tau_{circle}$ denotes the residence time in the one-shell configuration, respectively (5) or (6).
For the two selected systems these variations obey Kramers' relation.

According to this boltzmanian description, we will characterize in the following the configurations transitions by $R_J$ and by the mean residence times in each configuration.
These parameters depend on the  configuration  energy spectra and thus, are extremely sensitive to the local symmetry of the configurations.

\subsection{Lindemann-like criterion}

In order to explore the collective excitations of Wigner islands through a Lindemann criterion,
 the mean square deviations of the balls from their equilibrium locations have to be calculated.

For three-shell configurations and circular symmetry \cite{bedanov94}, we have to calculate, for each configuration and for shells 1 and 2, the radial displacements
$$u_r^2=\frac{1}{N_s}\sum_{i=1}^{N_s}(<r_i^2>-<r_i>^2)/r_0^2 ,$$
the relative angular intrashell  displacements
$$u_{\theta 1}^2=\frac{1}{N_s}\sum_{i=1}^{N_s}(<(\theta_i-\theta_{i_1})^2>-<\theta_i-\theta_{i_1}>^2)/\theta_0^2,$$
and, for each configuration and for shell 1, the relative angular intershell displacements
$$u_{\theta 2}^2=\frac{1}{N_s}\sum_{i=1}^{N_s}(<(\theta_i-\theta_{i_2})^2>-<\theta_i-\theta_{i_2}>^2)/\theta_0^2$$

where $(r_i,\theta_i)$ are the polar coordinates of a ball with respect to the center of the confining frame, $i_1$ indicates the right neighbor of the ball $i$, and $i_2$ indicates
its nearest neighbor in the surrounding shell, which is determined every snapshot. $r_0=1/\sqrt{\pi n}$, where $n$ is the balls density, is the mean
 radial free space for a ball and $\theta_0=2 \pi/N_s$ is the mean interball angular distance in the shell consisting in $N_s$ balls. $<>$ indicates an average over time.

These deviations are relative displacements (with respect to another ball or the center of the system), therefore they are relevant in order to exhibit a
Lindemann-like criterion comparable to the one used for larger two-dimensional systems \cite{bedanov85_2}. However, we will not be able to discuss as in the latter article about the
values of the dimensionless parameter $\Gamma=E/k_B T$, where $E$ is a typical interaction energy between two particles, since we do not have numerical
values for the potential energies in our system at the present time.

Note that we have discriminated between the different configurations. This is allowed since the radial displacements, as we shall show, are around a sixth of the distance between the two
 shells whatever the effective temperature, then the shells are always well identified.  This choice is
motivated by our requirements of precise information about the influence of the local order on the ``melting''. The temperature dependencies of the displacements averaged
over all the configurations have also been calculated. This procedure is not the same as the one used in \cite{bedanov94}, where
the distinction between shells is only made at the beginning of the numerical simulation, so we shall expect lower values for our radial displacements that do not include intershell jumps.

\section{\label{transition}Identification of transition temperatures}

     \begin{figure*}
\resizebox{1.9\columnwidth}{!}{
\includegraphics{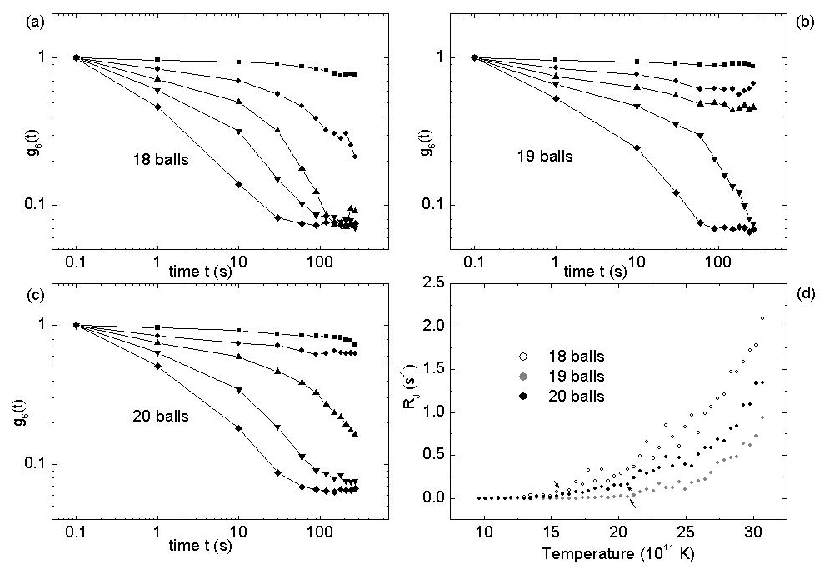}}
\caption{\label{correlationsauts}Variations of orientational correlation function and jump rate with temperature. Bond-orientational
 time correlation function for five temperatures : $(\blacksquare)\ T=9.6\times 10^{11} K$, $(\bullet)\ T=13.0\times 10^{11} K$,
$(\blacktriangle)\ T=17.8\times 10^{11} K$, $(\blacktriangledown)\ T=23.0\times 10^{11} K$, $(\blacklozenge)\ T=30.2\times 10^{11} K$ ;
 (a) $N=18$, (b) $N=19$ and (c) $N=20$. (d) Jump rate versus temperature. Arrows show the temperature at which the second excited state is reached.}
\end{figure*}

\subsection{Bond-orientational decorrelations}

On figure \ref{correlationsauts}(a-c) we present typical time correlation functions $g_6(t)$ for five different temperatures and for the three systems, for which the same behaviors are observed : at low
temperature, a constant value is reached, indicating an ordered state. At higher temperature, the correlations decay strongly. As in large systems, we can define a ``melting''
 temperature $T_{l}(N)$. We notice that the $N=19$ system keeps an ordered structure denoted
 by a quasi-constant correlation function until a much higher temperature than the two other systems. Indeed, the transition temperatures $T_{l}(N)$ are
sensitively different according to the number of balls $N$ : $T_{l}(18)$ is less than $13\times 10^{11} K$, whereas $T_{l}(19)$ is more than $18\times 10^{11} K$ and $T_{l}(20)$ between both.
Note our goal is not to measure precise transition
temperatures, but to put in evidence the mechanisms involved in the disordering. As we shall see, the different transition temperatures are sufficiently separated to do such an analysis.
 In the following,
we will then investigate the two expected mechanisms for this disordering : configuration transitions and collective excitations through Lindemann criterion.

\subsection{Configurations transition}

The first indication of the configuration transition activity can be evidenced by observing the evolution of the jump rate $R_J$ while the effective temperature increases.
These variations are shown on figure \ref{correlationsauts}(d) for the different studied systems.

Whatever the number of balls, $R_J$ presents the same qualitative behavior: it is close to zero at low temperature and increases
 strongly at higher temperature. These variations correspond to a progressive augmentation of the number of transitions activated at the effective
 temperature. Whatever the temperature, the smallest jump rate is associated to $N = 19$ system and the one associated to
$N=20$ is always smaller than the one corresponding to $N=18$. In order to describe more quantitatively these behaviors and their differences, let us
 introduce transition temperatures which characterize the ``beginning" of the $R_J$ increases. We have chosen to name transition temperature the
 temperature at which the ground configuration begin to switch. Let us nevertheless indicate that this transition temperatures does not correspond to an actual
sudden transition since the jump rate $R_J$ rises progressively. Note that experimentally, the infinite time limit in the $R_J$ definition corresponds to the maximum
measurement time and its finite value could alter the $R_J$ value. However a complete analysis has shown that the variation with temperature of $R_J$ is independent from this
 experimental limit provided that this limit was largely higher than the mean residence time of the system in each configuration. In our experiments we chose $t_{max}=400$s
 which satisfied this condition.

The analysis of the evolution of the residence times distribution shows that, for $N=19$, the transition temperature $T_{J}(19)$ is $20\times10^{11}$K, much higher than the temperatures $T_{J}(18)$ and $T_{J}(20)$
 which are respectively $14\times10^{11}$K and $17\times10^{11}$K. At larger temperature, the second excited states are
reached. The corresponding temperatures are respectively $21\times10^{11}$K, $16\times10^{11}$K and $21\times10^{11}$K for $N=19$, 18 and 20 and are indicated by arrows on
 figure \ref{correlationsauts}(d).  Those temperatures might correspond to the critical temperatures given in \cite{schweigert98_1} where the rate of radial
 jumps is calculated. This is exactly the same parameter as our jump rate since each configuration switch involves only one ball jump from one shell to another.

\begin{figure}

\resizebox{0.95\columnwidth}{!}{\includegraphics{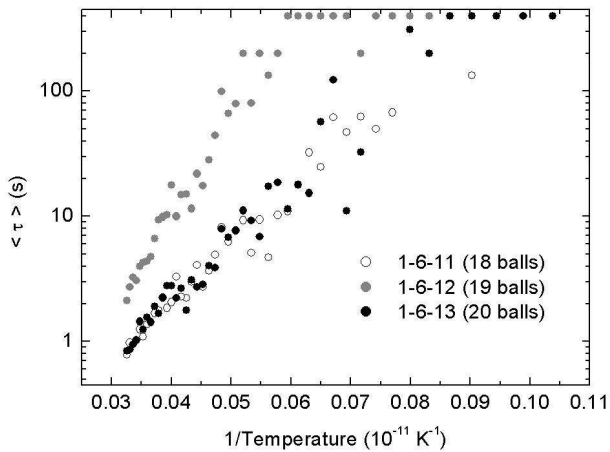}}
\caption{\label{tempsfond} Evolution with the temperature of the residence mean times in the ground state (the 400s limit  corresponds to the recording time
 but it is actually infinite). Times are in log scale.}

\end{figure}

These behavior differences between the systems are still more obvious when we study their mean residence times
 in the ground state. On figure \ref{tempsfond} we have plotted the logarithm of these times $<\tau>$ versus the inverse of the effective temperature.
We obtain a linear variation, which is in agreement with the Boltzmann law description presented above, the curves slopes being equal
 to the barrier heights for escaping the configuration. We can observe that these slopes are different according to the number of balls. The highest slope is associated to
 the ground state $N=19$ and is equal to $192\times10^{11}$ K, those corresponding to $N=18$ and 20 being respectively equal to $99\times10^{11}$K
and  $107\times 10^{11}$K.
 This indicates that the ground state for $N= 19$ is much more stable than the ground states associated to $N=18$ and 20, whose barrier heights are very similar
since their corresponding slopes are roughly identical.

\begin{figure}
\resizebox{0.95\columnwidth}{!}{
\includegraphics{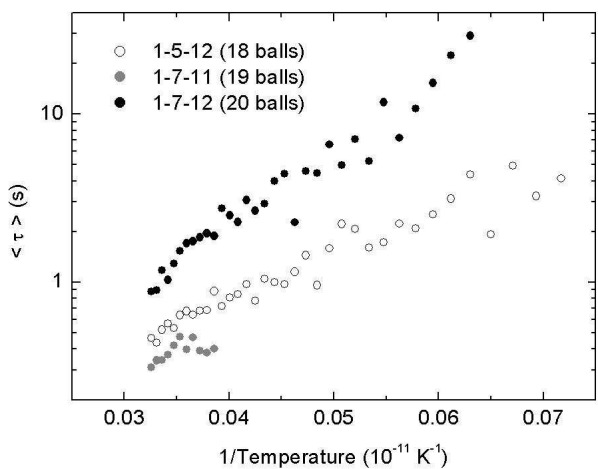}}
\caption{\label{tempsexc}  Evolution with the temperature of the residence mean times in the first excited state. Times are in log scale.}

\end{figure}

This analysis can be completed by studying the excited states (figure \ref{tempsexc}).
The slopes for $N=18$  is $64\times10^{11}$K and is equal to $101\times10^{11}$K for $N=19$ ; in both systems, this suggests that the barrier is much lower than the ground state's one.
 On the contrary, the slope associated to the first excited state of $N=20$ is equal
 to $96\times10^{11}$K, almost the same as the barrier for the ground state. All these results are in good agreement with the energy levels calculated in \cite{campbell79}.
Let us indicate that even when they are reachable,
 the other excited states are not easy to study since their higher energy involves too small statistical occurrences. Finally, these measurements allow us to
determine the first excitation energy associated to each system. These energies are $35\times10^{11}$K, $91\times10^{11}$K, $11\times 10^{11}$K respectively for $N= 18$, 19, 20.

These measurements show that the ``magic number "system corresponds to the deeper ground
 state and confirms its strong stability in comparison with the two other systems. As expected, the commensurability influences
strongly the depth of the well, and consequently the transition temperature in $R_J$.

Let us conclude this section by indicating that obviously the jump rate $R_J$ is related to the mean residence times. For instance, for a two-level system, $R_J$ is
 simply equal to $2/(<\tau_1>+<\tau_2>)$, where the subscripts 1 and 2 stand for the two levels. This relation is satisfied and confirms the self consistency
of our results, at least up to a temperature at which a third level is reached.

\subsection{Mean square displacements}

We now turn to the study of the ``melting'' through Lindemann-like criterion. We will successively present the radial, intrashell and intershell mean
square displacements averaged over all the configurations. In order to explore more precisely the relation between the local order and these collective parameters,
 we have compared them to the corresponding parameters before averaging. We shall show in particular that the procedure of averaging over the configurations,
 commonly used in literature, mask actually subtle effects resulting from the configurations transitions, even though they don't infer that much on transition temperatures.

\subsubsection{Radial displacements}

\begin{figure*}
\resizebox{1.9\columnwidth}{!}{
\includegraphics{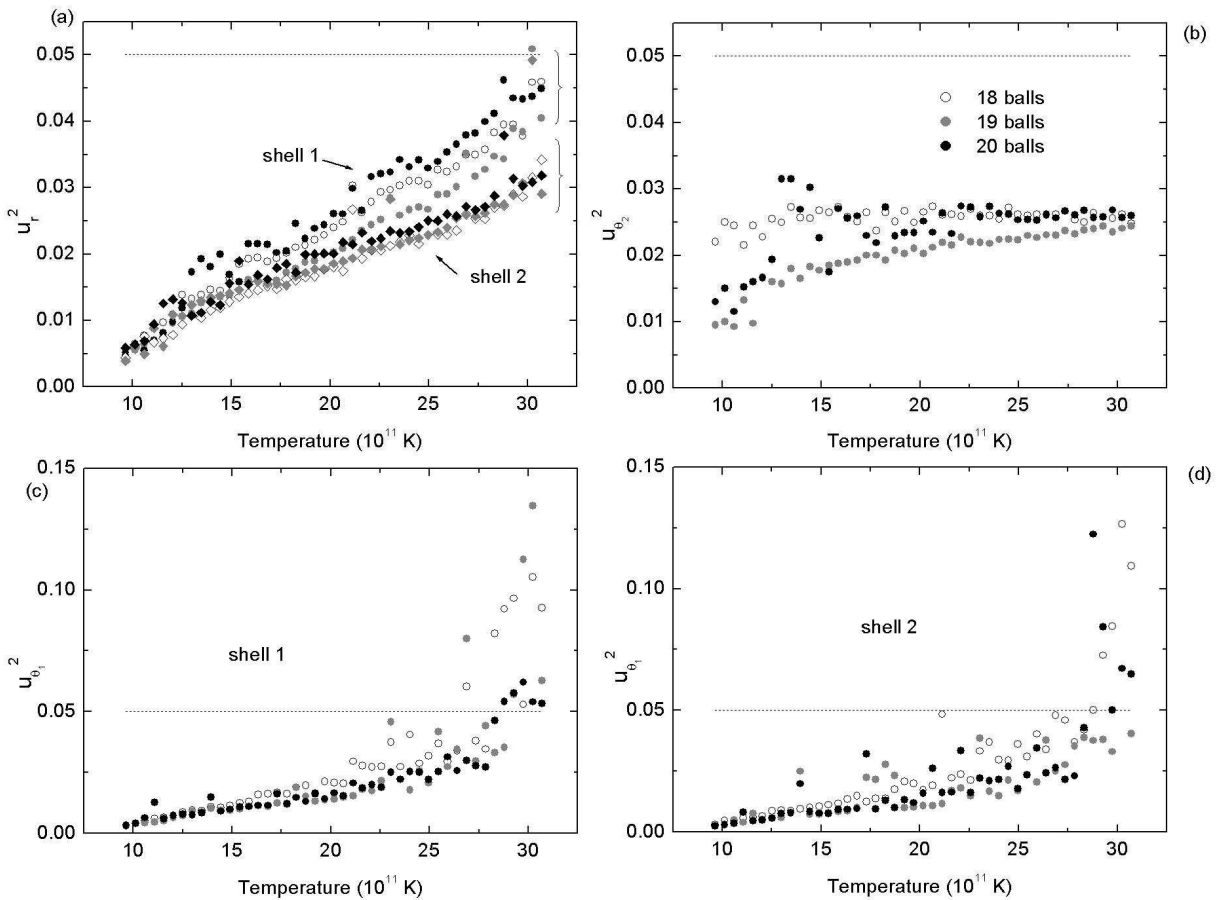}}
\caption{\label{exc} Mean square displacements averaged over the three configurations versus temperature : (a) radial displacements $u_r^2$ for both shells, (b) intershell angular displacements $u_{\theta 2}^2$,
(c) intrashell angular displacements $u_{\theta 1}^2$ in shell 1 and (d) intrashell angular displacements $u_{\theta 1}^2$ in shell 2}

\end{figure*}

The temperature dependencies of the radial mean displacements averaged over all the configurations are presented for the three different numbers of balls on figure \ref{exc}(a).

The displacements corresponding to the inner shell (shell 1) present similar behaviors for the three systems. They vary regularly from 0.005 at low temperature
and go on increasing until 0.05 at the highest experimental temperature, the highest and the lowest radial displacements being respectively associated to the systems $N=20$ and $N=19$,
whatever the temperature. In this experimental temperature range, the low values of $u_r^2$, smaller than the Lindemann criterion, indicate that each shell remains
 very well identified and that we can always discuss qualitatively the results in terms of shell entities.
 Notice that the highest obtained temperature corresponds to the ball-tracking limit, but we could forecast that the value would increase strongly beyond 0.05 after this temperature limit, since this value is far from being the maximal possible value, even though jumps from one shell to another are not taken into account.
 From this point of view, we can consider the
 highest experimental temperature is very close to the radial transition temperature transition $T_R$ of these systems.
The radial displacements of the outer shell (shell 2) are qualitatively similar to those observed for the shell 1 although their values are smaller and
 vary from 0.005 until 0.03.  This can be explained by the fact that this shell is submitted to a regular
and constant potential resulting from the confinement frame whereas shell 1 is submitted to fluctuant potential from both sides \cite{bedanov94}.

\begin{figure*}
\resizebox{1.9\columnwidth}{!}{
\includegraphics{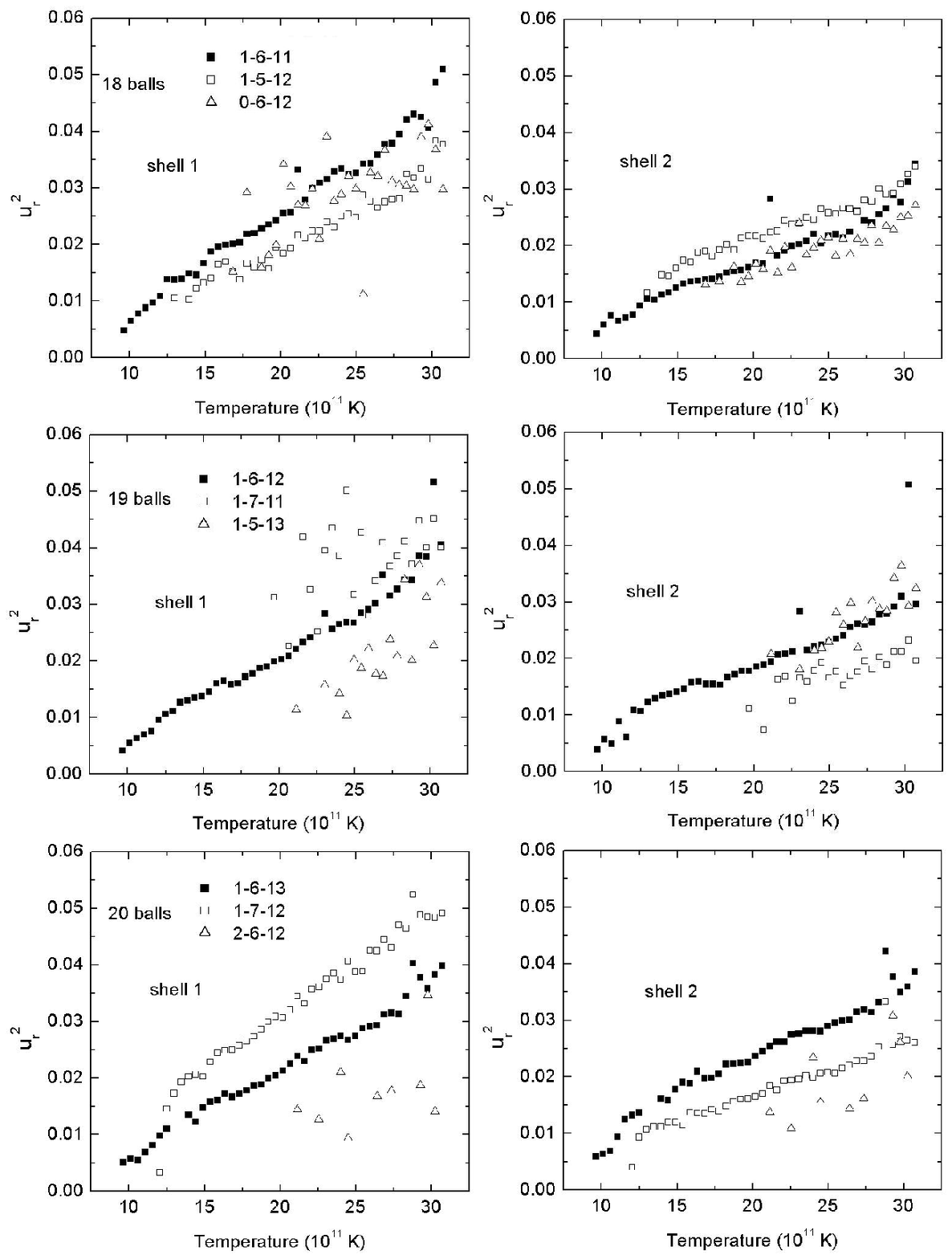}}
\caption{\label{excrad} Radial mean square displacements $u_r^2$ for the three systems and all configurations versus temperature.}

\end{figure*}

We have seen that  geometrical considerations are responsible for the different configuration transition rates observed.
 It is therefore natural to examine if it is the same for the displacements. The
 radial mean displacements for the different configurations of the different systems are presented on figure \ref{excrad}.
Notice that data for the excited states with low residence time present higher statistical error due to the smaller number of their occurrences;
 this is the case for instance for the second excited state for 18 and 20 balls and for both excited states for 19 balls.

As for the averaged curves, the radial displacements in the two shells of the systems in their ground configurations present regular increases with temperature.
 More precisely, for shell 1, we can observe that the $N=19$ (1/6/12) and $N=20$ (1/6/13) curves are identical whereas the displacement
 corresponding to $N=18$ (1-6-11) is higher whatever the temperature. For shell 2 the displacements  are identical for $N=18$ (1/6/11) and $N=19$ (1/6/12)
 whereas those associated to $N=20$ (1-6-13) presents a higher value. The temperature dependencies observed for the excited states
look like those associated to the ground states, with a constant switch. The highest $u_r^2$ values are now observed for $N=20$ (1-7-12) in the case of the
shell 1 and for $N=18$ (1-5-12) in the case of the shell 2.

 \begin{figure*}
\resizebox{1.9\columnwidth}{!}{
\includegraphics{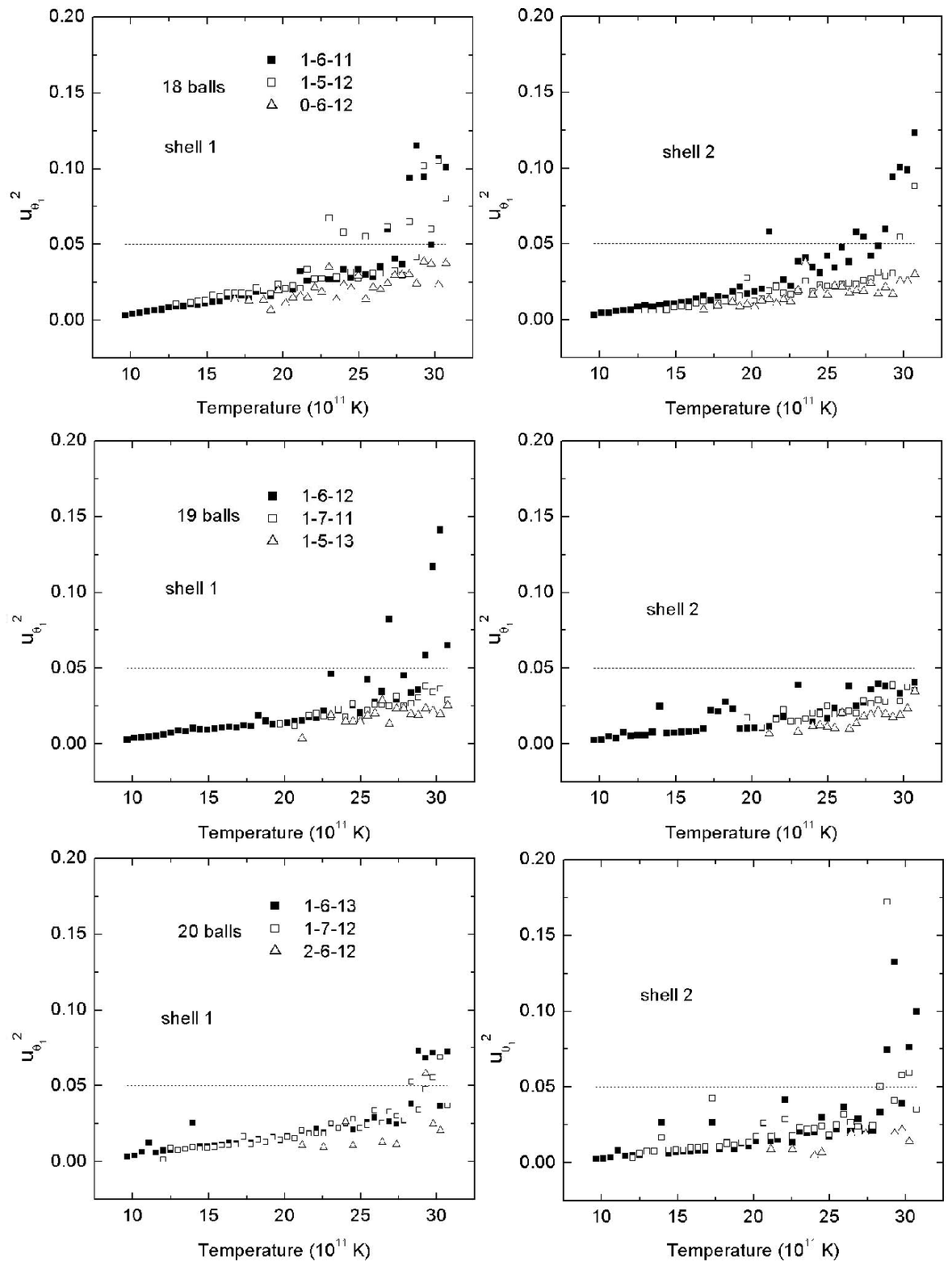}}
\caption{\label{excort} Intrashell mean square displacements $u_{\theta 1}^2$ for the three systems and all configurations versus temperature.}

\end{figure*}

We cannot define a precise rule to explain the different behaviors. It seems however that, for a given temperature, the more balls in the shell, the larger the radial displacements.
In the case of equality (that's to say, for the first shell, the couples (1-6-11),(0-6-12) and (1-6-13),(2-6-12) and
for the second shell the couples (1-5-12),(0-6-12) and (1-7-12),(2-6-12)), we observe that the commensurable states have lower radial displacements. Those differences are higher than a possible
 distortion due to a renormalization by a different $N_s$.

The first statement can be simply understood by the fact that orthoradialy squeezed balls have more kinetic energy to involve in their radial movement. In the case of a
commensurable state, all the balls (for the first shell) or half of them (for the second shell) find themselves in front of a repelling ball of the other shell, which restrains their radial fluctuations. Note this result is opposite to what
 was reported for quantum dots in \cite{filinov01} : the authors exhibit higher radial displacements for magic number $N=19$.

Beyond these specific results, the comparison between the  averaged and non averaged radial displacements show that even if
some differences can be exhibited concerning their precise and relative values, global behavior remains the same : each shell will radialy melt at the same temperature whatever the configuration.

\subsubsection{Intrashell angular displacements}

In the case of the angular mean displacements, the averaging does not introduce any distortion in the analysis, as it can be seen on figure \ref{excort}. So we will only discuss
 the mean displacements averaged
over the configurations (figures \ref{exc}(c) and (d)). Like the radial displacements, the intrashell orthoradial displacements associated with the inner shell of ground configurations
increase with the temperature.
However, whereas the former keep on rising slowly, the angular ones begin to vary linearly and change very rapidly at almost the same temperature whatever the number of balls.
By contrast, this similarity of behaviors is not observed for the outer shell  : whereas the intra-shell orthoradial displacements are identical for $N=18$ and $N=20$, in the case $N=19$,
it remains smaller than 0.005 and without rapid rise.

The changes are observed when the displacements reach the critical value $0.05$ in accordance with Lindemann criterion. Thus, we can define from these data
a transition temperature $T_O$ or, rather, a temperature interval centered temperature on $T_O$ (between $28\times10^{11}$K and $30\times10^{11}$K) after which intra shell orientational order
is lost in a given configuration for $N=18$ and $N=20$. We can expect that the corresponding temperature is not far from the experimental limit in the case of $N=19$ since the balls in the shell 1 have begun
 to be non-correlated whereas the shell 2 is still rigid.

\subsubsection{Intershell angular displacements}

Let us now consider the intershell relative displacements which measures the ability of the two shells to find a stable position one with respect to the other. Their variations with temperature
 averaged over configurations
 are presented in figure \ref{exc}(b) for the three systems.
We can notice that even at low temperature these displacements are high and until $T\approx18\times10^{11}$K, the intershell angular displacements are larger than the corresponding
 intrashell and radial displacements. This indicates that at low
 temperature these intershell movements are the main effects resulting from the thermal fluctuations. Moreover, this temperature being smaller than $T_{O}$ and $T_R$, we are allowed
to discuss these results in terms of well defined rigid shells in which the balls are regularly spaced.

At large temperature, $T> 18 \times10^{11}$K, the inter-shell angular displacements reach the same finite value whatever the system, this value being in fact its theoretical maximum. In
this temperature range, the thermal energy is sufficient to overcome the barrier energies, which correlate the inner and outer shells. Since disorder is characterized by a deviation of the
whole island from the symmetrical situation, note that the intershell displacements are calculated considering the nearest neighbor at each time step, in order to
take into account the invariance towards some rotations. Consequently, we will always have a small maximum value and will never observe a steep rise as in \cite{bedanov94}.

Below this temperature, a temperature range in which the effect of the local order has an essential play is well exhibited. In this temperature domain, the systems are mainly
 in their ground state and the intershell angular
 displacement is lower for 19 balls than for 18 and 20 balls, the latter remaining of the same order. This can be simply explained by commensurability
 arguments : let us consider the shells as rigid rings ;  in the 1-6-12 configuration, shell 1 is submitted to a $2 \pi /12$ periodic
 potential due to shell 2, then each of its balls can find itself in a potential well. On the contrary, in any other configuration uncommensurability implies that, if rigid,
 the two shells can't find a position for which every ball will be located in a minimum of energy, hence a higher instability.

\section{\label{order} Disordering scenario}

 \begin{figure*}
\resizebox{1.9\columnwidth}{!}{
\includegraphics{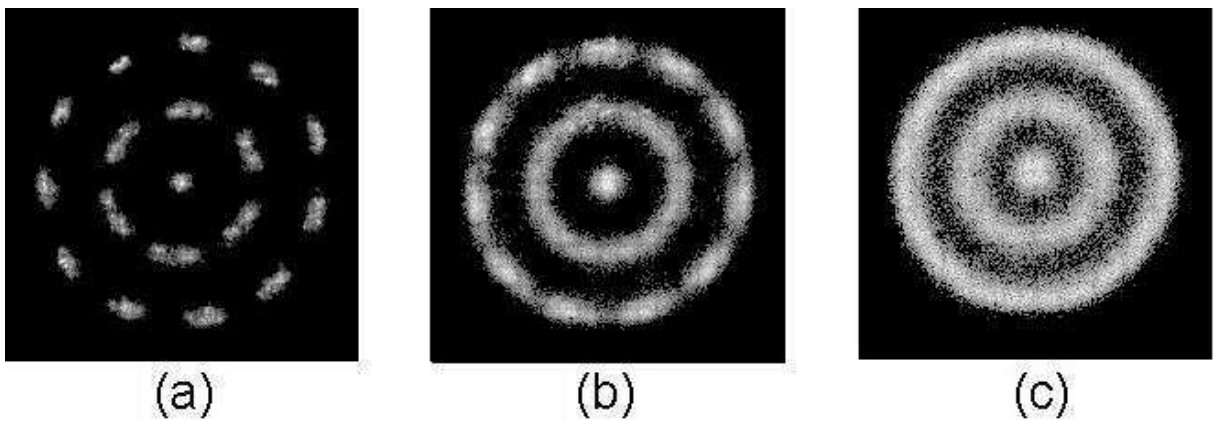}}
\caption{\label{cm} Mean crystals for (a) $T<T_{J}$, (b) $T_O<T<T_{R}$ and (c) $T\sim T_{R}$ .}

\end{figure*}
The study of the transition temperatures $T_l$ through correlation functions show that they depend strongly on the local order. Indeed, those temperatures
 are very different for the three systems, the $N=19$ ``magic system'' being more stable.
As for temperatures around $T_l$ , the systems can still be seen as sets of shells, comparison can be made between the temperatures $T_{l}$ and the temperatures $T_{J}$ of
first configuration transition. Since they are very similar, we can infer that configuration transitions play an important role in the disordering, at least for intermediate temperatures. It can be
simply understood by the fact that the islands are invariant by many rotations whatever the configuration, then a cycle of configuration transitions can induce
disorder (from the correlation point of view),  since the initial particles positions are not necessary recovered at the end of the transition cycle. Moreover a configuration
switch implies a complete reorganization in the shells. This disorder mechanism involving only one particle jump
cannot be evidenced by the global parameters as the mean displacements used in the Lindemann approach.
Note that  even at low temperature, the orientational order between
adjacent shells is already lost while their internal order is conserved ; this suggests that the trajectory of the jumping particle might be not only radial
 but also orthoradial, taking advantage of the relative rotation of the two implied shells. This is also in accordance with the fact that commensurability
considerations play a role for the intershell rotations as well as for the height of the energetical barriers. On the other hand, mere intershell rotation is not sufficient
 to induce disorder as defined by the correlation function, since the system will periodically
find itself in the same position as the initial one.

If we focus now on collective displacements, a first transition occurs at the temperature $T_O>T_l$,
 which corresponds to the emergence of the angular intra-shell diffusion. This stage in which the systems can
be considered as independent shells remains until a second transition temperature $T_R$ involving diffusion of particles between the shells.
At higher temperatures, the shell structure disappears. Figure \ref{cm} shows the mean crystals (superposition of the different positions of the particles) obtained for instance for $N=18$
at three key temperatures. Figure \ref{cm}(b) illustrates
the role of the transition activity in the loss of the orientational order : according to the values of the angular displacements, particles positions should be distinguishable.

This kind of two-stage ``melting'' had been suggested by previously cited works, especially in  \cite{bedanov94}, where the authors
focus on a global Lindemann criterion that includes different mecanisms that we have exhibited there. In particular, we have distinguished two contributions in the radial displacements, namely
 the individual jumps and the mean behavior. In colloidal systems \cite{bubeck99}\cite{bubeck02}  the system is arranged at low temperature in a shell-like
 structure. It also exhibits a very similar behavior when the temperature increases, excepted a re-entrant ordered phase which
was not observed in our case, this phase being specific to the hard wall confinement \cite{schweigert00}.

In \cite{filinov01}, the authors studied the melting of $N=19$ and $N=20$ quantum particles
interacting with coulombic interaction. They described the melting
as a two stages process :  first an orientational inter and intrashell disordering and then
a radial melting at higher temperatures. The relative positions of the transition temperatures found here are in good qualitative agreement with their results.
In addition, we can clearly distinguish two phases in the angular
disordering, namely an intershell rotation and then an intrashell melting. This last point is also presented in numerical works on classical coulombic particles  \cite{bedanov85_2}.

Whereas local geometry  have an influence on the configuration transitions through commensurability considerations, we
 have shown that its effects are neglectable for the intrashell displacements as well as for the radial ones.  On the other hand, correlation
 functions define very different temperatures for our three systems (all the transition temperatures are summarized in table \ref{table}). Moreover, and contrary to large systems, those
temperatures are lower than the transition temperatures given by the Lindemann criterion.
We have then to consider that beyond the
well-known Lindemann scenario, there are other sources of disorder, namely the configuration switches.

 \begin{table}

\begin{tabular}{ccccc}
N&$T_l$&$T_J$&$T_O$&$T_R$\\
\hline
\hline
18&$T_l<13$&14&29&$T_R>30$\\
19&$T_l>18$&20&30&$T_R>30$\\
20&$13<T_l<18$&17&29&$T_R>30$\\

\end{tabular}

\caption{\label{table}Summary of the different transition temperatures (in $10^{11}K$).}

\end{table}

\section{Conclusion}

In this paper we show that for a system constituted of a small number of interacting particles like Wigner islands, an increase of temperature results in  a disordering of the
 system more than a real melting.  This disordering process is very sensitive to the
local order of the explored configurations.

This disordering results from both individual excitations that induce configuration
transitions and collective excitations. This process is marked by three different transitions. The temporal
correlation functions which describe the correlation loss of the system exhibits an exponential decrease at the
"liquid transition temperature" $T_l$ (named after the usual convention). This first transition is identified without
ambiguity as corresponding to the increase of configuration transitions between the stable and metastable states of the system.
 So the liquid transition temperature depends strongly on the local order as the transition rate. At larger temperatures, two other transitions appear,
$T_O$ and $T_R$ characterizing respectively intrashell and intershell diffusion. These transitions are evidenced by the change of the mean square
displacements with temperature and correspond to collective excitations. The local order play a less significant role in the transition temperature values.

This disordering process and the importance of the local order on these temperatures are due the small
number of configuration states explored by the system at a fixed temperature. From this point of view, it is a specific
characteristic of small systems. Indeed, these effects cannot be observed for
large systems since the number of explored configurations is large enough to mask the individual excitations in the collective ones,
hence the coincidence between the transition temperatures described by the correlation functions and those identified by the Lindemann criterion.

Whereas the $T_l$ transition was never discussed before, the two following successive transitions have previously been
mentioned in the literature. The corresponding analyses are in agreement with our results but their approach
 is very different. We show that the description of the transition from well organized arrays towards liquid state resulting from successive
excitations requires more detailed analyses than the single use of Lindemann criterion.

Finally, let us conclude by suggesting that small Wigner islands that we proposed could be good candidates in order to
 explore experimentally the thermodynamic laws dedicated to small systems of interacting particles\cite{smallsys}.

\end {document}